\documentclass[a4paper]{article}

\usepackage{arxiv}

\usepackage[utf8]{inputenc} 
\usepackage[T1]{fontenc}    
\usepackage{xcolor}
\definecolor{darkred}{rgb}{0.65,0.0,0.0}
\definecolor{darkblue}{rgb}{0.0,0.0,0.5}
\usepackage{hyperref}       
\hypersetup{                
    colorlinks=true,
    linkcolor=darkblue,
    filecolor=darkblue,
    urlcolor=darkblue,
    citecolor=darkblue
    }
    
\usepackage{url}            
\usepackage{booktabs}       
\usepackage{amsfonts}       
\usepackage{nicefrac}       
\usepackage{microtype}      
\usepackage{lipsum}		
\usepackage{graphicx}
\usepackage[numbers]{natbib}
\usepackage{doi}

\usepackage{amsmath}
\usepackage{siunitx}
\usepackage{caption}
\usepackage{subcaption}
\usepackage{float}
\usepackage[inkscapeformat=png]{svg}
\usepackage{tikz}

\usepackage{tabularx}
\usepackage{listings}
\usepackage[section]{placeins}
\DeclareUnicodeCharacter{2212}{-}

\usepackage{xcolor}
\definecolor{codegreen}{rgb}{0,0.6,0}
\definecolor{codegray}{rgb}{0.5,0.5,0.5}
\definecolor{codepurple}{rgb}{0.58,0,0.82}
\definecolor{backcolour}{rgb}{1,1,1}

\lstdefinestyle{code_style}{
    backgroundcolor=\color{backcolour},   
    commentstyle=\color{codegreen},
    keywordstyle=\color{magenta},
    numberstyle=\tiny\color{codegray},
    stringstyle=\color{codepurple},
    basicstyle=\ttfamily\footnotesize,
    breakatwhitespace=false,         
    breaklines=true,                 
    captionpos=b,                    
    keepspaces=true,                 
    numbers=left,                    
    numbersep=5pt,                  
    showspaces=false,                
    showstringspaces=false,
    showtabs=false,                  
    tabsize=2
}

\usepackage{authblk}

\title{A waypoint based approach to visibility in performance based fire safety design}

\author[a]{{Kristian Börger\hspace{0.1em}}}
\author[b]{{Alexander Belt\hspace{0.1em}}}
\author[a,b]{{Lukas Arnold\hspace{0.1em}}}

\affil[a]{\hspace{0.1em}Chair of Computational Civil Engineering, University of Wuppertal, Pauluskirchstraße 7, 42285 Wuppertal, Germany}
\affil[b]{\hspace{0.1em}Institute for Advanced Simulation, Forschungszentrum  J\"ulich, Wilhelm-Johnen-Straße, 52428 J\"ulich, Germany\newline}

\affil[ ]{
Kristian Börger: \texttt{\href{mailto:boerger@uni-wuppertal.de}{boerger@uni-wuppertal.de}; \href{https://orcid.org/0000-0003-4371-5770}{ORCID: 0000-0003-4371-5770}}
}

\affil[ ]{
Alexander Belt: \texttt{\href{mailto:a.belt@fz-juelich.de}{a.belt@fz-juelich.de}; \href{https://orcid.org/0000-0002-6091-9321}{ORCID: 0000-0002-6091-9321}}
}

\affil[ ]{
Lukas Arnold: \texttt{\href{mailto:l.arnold@fz-juelich.de}{l.arnold@fz-juelich.de}; \href{mailto:arnold@uni-wuppertal.de}{arnold@uni-wuppertal.de}; \href{https://orcid.org/0000-0002-5939-8995}{ORCID: 0000-0002-5939-8995}}
}

\renewcommand{\shorttitle}{\textit{A waypoint based approach to visibility in performance based fire safety design}}

\hypersetup{
pdftitle={A waypoint based approach to visibility in performance based fire safety design},
pdfsubject={A waypoint based approach to visibility in performance based fire safety design},
pdfauthor={Kristian Börger, Alexander Belt and Lukas Arnold},
pdfkeywords={FDSVismap; visibility map; visibility; FDS; smoke modelling; extinction coefficient; performance based design; ASET-RSET concept},
}

\begin{document}
\maketitle

\begin{abstract}

In performance-based fire safety design, ensuring safe egress, e.g. by visibility of safety signs, is a crucial safety goal. Compliance with the building requirements is often demonstrated by simulations of smoke spread. Numerical models like the Fire Dynamics Simulator generally compute visibility as a local quantity using the light extinction coefficient, without the consideration of the actual light path to a safety sign. 
Here, visibility maps are introduced, providing an approach for post-processing fire simulation data. They indicate safe areas along egress routes, with respect to visibility. At each location, the available visibility is calculated using Jin's law, as an integrated value of the extinction coefficient along the line of sight to the closest exit sign. The required visibility results from the distance between those points. Additional parameters like view angle or visual obstructions are considered. The presented method allows for temporal visibility assessment, e.g. in an ASET-RSET analysis.

\end{abstract}

\keywords{FDSVismap \and Visibility map \and Visibility \and Fire Dynamics Simulator \and Computational Fluid Dynamics \and Smoke Modelling \and Extinction Coefficient \and Performance Based Design \and ASET-RSET Concept}

\section{Introduction}
\label{sec:introduction}

Safety signs play a significant role in protecting building occupants by guiding them in the event of an emergency. Design principles of safety signs, incorporating dimensions and colorimetric and photometric characteristics, are internationally standardised in ISO 3864-1 \cite{din_din_iso_3864_1}. Additionally, ISO 7010 \cite{iso_iso_7010} specifies certain safety signs, including exit signs. Exit signs are intended to ensure a safe egress by clearly indicating the location of the closest emergency exit. Hence, extensive research has been conducted in the past on various aspects that physically or physiologically affect the perception or interpretation of exit signs.

Chen et al. proposed an algorithm to optimise the arrangement of exit signs in public buildings in order to reduce the overall evacuation time \cite{chen_location_2009}. They address the issue of most building codes not providing detailed guidelines or clear standards for placing exit signs. The algorithm is intended to maximise the coverage of the signs and enhance the evacuation efficiency based on a cellular automaton evacuation model. In an experimental study, Shi et al. investigated the influence of different colours and types of exit signs on the attention and perceptual processing \cite{shi_influence_2022}. They conclude that the participants had a longer processing time for red than for green signs. Green signs on the other were causing stronger attention bias, which is an indicator for the difference in assigning attention to a certain stimulus when being faced with different stimuli.

The perception of exit signs in case of a smoke induced impairment of visibility was first comprehensively investigated by Jin in the 1970s \cite{Jin_1970}. From the visual obscuration threshold at which the sign can still be seen, he derived an empirical law on the correlation of the smoke's light extinction coefficient $\sigma$ and a visible distance (visibility) $V$. With constant boundary conditions of smoke density, smoke characteristics and ambient lightning, all factors affecting the contrast can be encapsulated by a single visibility factor $C$, simplifying this correlation to Eq.~\ref{eq:jins_law}.

\begin{equation}
    V = \frac{C}{\sigma}
    \label{eq:jins_law}
\end{equation}

For white smoke, as employed in his experiments, he found $C$ to be $2\sim4$ for light reflecting signs and $5\sim10$ for light emitting signs. In subsequent studies, Jin examined the validity of this correlation for different types of smoke resulting from flaming combustion (black smoke) or smouldering (white smoke). He observed that at the same smoke density, black smoke allows for a greater visibility than white smoke \cite{Jin_1971}. Further research involving an actual exposure of participants with smoke revealed that visibility is significantly reduced by eye irritating effects when the smoke density exceeds a certain threshold. Impaired vision can result in psychological and physiological effects, including panic and a decrease in walking speed \cite{Jin_1972}.

In another study, Jin conducted measurements of the extinction coefficient at different wavelengths. At the beginning of the experiments, the extinction coefficient decreases with a higher wavelength in the visible range of light. With a dependence on the soot particle size distribution, the effect can change and even reverse with time due to the ageing effects of the aerosol. From these insights, he derived that the visibility for red and blue signs differs: the visibility for red signs within smoke with predominantly scattering properties (white smoke) is 20 - 40\% higher, and within predominantly absorbing smoke, it is 20 - 30\% higher. These conclusions have been drawn by directly correlating the extinction coefficient with visibility, rather than conducting separate investigations. The underlying assumption is that both signs have identical brightness and obscuration thresholds to be perceived. A different approach to assess visibility as a function of exit signs' colour was followed by Oh et al. \cite{oh_evaluation_2023}. They simulated a smoke-filled environment by having study participants wearing translucent eye patches featuring nine different levels of visual obstruction. Contrasting to Jin's conclusions, the visible distance at which exit signs were still detectable was significantly higher for blue signs than for red and green signs. Although the experiments did not take into account the wavelength-dependent characteristics of the fire smoke, a significant effect of colour on the perceived contrast and thus visibility can be drawn from their observations.

Jin's law is still widely used to assess visibility in performance-based fire safety design. It also serves as a standard model in numerical fire models such as the Fire Dynamics Simulator (FDS). The model allows a temporally and spatially resolved output of the local visibility, as a function of the respective extinction coefficient. Awadallah et al. state that the interpretation of visibility on simulation data such as horizontal slices of the computational domain can be highly subjective when evaluated by different engineers \cite{awadallah_nfsd_2022}. They also note that Jin's analytical model was originally developed for the assessment of visibility in homogeneous smoke and should be used carefully on non-uniform smoke environments. Hence, they propose a numerical approach for averaging in the evaluation of visibility. The concept of an averaged extinction coefficient is also applied by Rinne et al. to evaluate the visibility of exit signs through an inhomogeneous smoke layer based on FDS simulation data \cite{2007.Rinne}. Węgrzyński et al. propose a similar approach involving ray tracing for a more realistic assessment of visibility in the context of CFD (Computational Fluid Dynamics) simulations \cite{wegrzynski_experimental_2017}.

Husted et al. presented a model that allows considering an inhomogeneous smoke distribution when assessing visibility based on CFD simulations in the context of performance-based design \cite{husted_bjarne_paulsen_visibility_2004}. By applying a procedure for averaging the smoke density, their model allows to evaluate visibility with respect to the view point and viewing direction. Compared to a local or cell-based assessment of visibility, the approach allows indicating more realistically if an object would be recognisable in a real fire event. However, the model predominantly aims at the assessment of visibility in concise, single compartment geometries and thus may be unsuitable for evaluating escape scenarios in complex building structures.

In this study, an innovative method is presented to assess visibility in performance based design, utilising data from numerical fire models. The approach is designed for ease of application, providing a clear and realistic interpretation of visibility, even in complex building designs. It adopts Jin's law for scenarios with inhomogeneous smoke distributions, as derived from CFD simulation results. A waypoint-based approach is introduced to evaluate visibility along the entire route of egress. These waypoints, representing exit signs, serve as key reference points for computing the maximum visual distance through fire smoke from the surrounding areas. A simple ray-casting algorithm is applied to compute the averaged extinction coefficient along the line of sight.  Finally, this process results in the creation of visibility maps. Visibility maps describe a two-dimensional matrix of binary values, providing a localised indication of whether the next exit sign is visible at every point along the route of egress.

Visibility maps enable a more realistic and distinct assessment of egress in case of fire, considering smoke induced reduction of visibility. No performance criteria need to be defined due to the automated consideration of the building's architectural characteristics. Consequently, the personal bias in the interpretation of simulation results is significantly reduced, resulting in increased credibility in the approval process for building designs.

The model has been implemented as a Python package and is made freely available. Operating as a post-processor for results data from numerical fire simulations, the package can generally process data from any CFD model featuring a cubic mesh-grid discretisation. However, automatic data extraction is currently limited to FDS results, as it is based on the FDSReader \cite{fds_reader}.

\section{Visibility in performance based design}
\label{sec:visibility_in_performance_based_design}

Performance based design provides an alternative approach to applying prescriptive building regulations within the process of building approval. The SFPE Engineering Guide to Performance-Based Fire Protection \cite{sfpe_nfpa_performance_based_design_2007} and the Interpretative Document Fire Safety of the European Commission \cite{eu_interpreatative_document_1994} follow a similar definition of performance based design in fire safety as the engineering approach to satisfy fire safety goals and objectives. The primary safety goals of most national building codes, such as NFPA 101 \cite{nfpa_nfpa_101_2021}, address the health of building occupants while ensuring safe egress in hazardous events. While fire safety goals usually represent fundamental requirements, they are specified by means of objectives to be met. This might include the structural integrity of the route of egress, limiting the occupants' exposure to toxic smoke products, or ensuring sufficient visibility in a smoke laden environment.

Prescriptive regulations typically require designated architectural, structural, technical, or operational measures to meet these objectives. Especially in modern buildings, characterised by a complex or open architectural designs, meeting these requirements can often be challenging or involve disproportionate efforts. Therefore, methods of performance-based design are increasingly being utilised in fire safety engineering, to meet the specific needs of the building and its occupants. In this context, the objectives of the building regulations to be met are quantified as performance criteria that provide a threshold for compliance. Defining the objectives and performance criteria typically takes place within the approval process, in consultation with the authorities in charge. Several national building codes, such as Switzerland~\cite{VKF_nachweisverfahren_2017}, New Zealand \cite{dbh_nz_buildingcode_2023}, or the USA \cite{nfpa_nfpa_101_2021}, natively implement performance-based design and offer compliance with specific performance criteria as an alternative to satisfying prescriptive requirements.

In performance based fire safety design, evaluating a building design usually involves hazard calculations and fire dynamics calculations based on a predefined design fire scenario \cite{springer_sfpe_hurley_2016}. According to ISO 13943 \cite{iso_iso_13943} a fire scenario is defined as a ``qualitative description of the course of a fire with respect to time, identifying key events that characterise the studied fire and differentiate it from other possible fires''. From the chosen design fire scenario, the design fire can be derived, which is a ``quantitative description of assumed fire characteristics''. It typically reflects the temporal variation of key variables such as heat release rate (HRR), smoke production rate and the yields of toxic smoke products. Selecting relevant design scenarios from all potential scenarios may be conducted by either a deterministic or probabilistic approach. In contrast to consider ``worst-case'' scenarios, it should be aimed to select credible yet conservative scenarios, associated with the highest risk. This helps excluding scenarios that are either highly improbable or entail negligible consequences. This process may employ fully quantitative methods according to ISO 16732-1 \cite{iso_iso_16732_1} or qualitative or semi-quantitative approaches according to ISO 16733-1\cite{iso_iso_16733_1}.

Significant uncertainties in the context of performance-based design result from the input variables of fire models, in particular from the defined design fire scenario. Johanssen et al. have shown in a round-robin study among fire safety engineers that significant discrepancies can occur in modelling a fire in FDS, even when the scenario is described in a high level of detail. Besides different assumptions, mistakes, insufficient knowledge about the software, they identified insufficient knowledge about fire dynamics as the biggest source of uncertainty \cite{johansson_variation_2018}.

Visibility in case of fire constitutes a central objective to ensure successful egress of buildings occupants. It depends on the irritant effect and the smoke density, having physiological and psychological effects on the evacuees. This inherently affects both their walking speed and the ability of way finding \cite{nfpa_sfpe_jin_2002}. For buildings with high ceilings, like industrial facilities, the proof of save egress can usually be provided by estimating the height of a low-smoke layer above the floor level. This simultaneously limits both excessive thermal stress and people's exposure to toxic smoke products. However, such evidence may be complicated to provide, especially in buildings with complex geometries, e.g. featuring a multi-storey spanning open design. In such cases, evaluating the actual impairment of visibility due to fire smoke may be advisable.

Proof of save egress due to visibility based on numerical fire simulation models is usually conducted at a temporal level. In this context, the available time after the outbreak of the fire is determined until a route of egress is no longer passable and thus self-rescue is no longer possible. The evidence is provided if the available safe egress time (ASET) is greater than the required safe egress time (RSET) \cite{Cooper_asetrset_1983}. RSET can be specified as a global performance criterion or individually derived from an evacuation simulation, as outlined in \cite{schroder_map_2020}.

Various tenability criteria for the minimum visibility required for a successful escape can be found in the literature. Those are subject to considerable scattering and depend largely on both the buildings and the occupants' characteristics. Short distances to emergency exits, for example, may require less stringent thresholds if the occupants are familiar with the building. Negative influences include poor lighting conditions, complex geometries, and expansive spaces~\cite{nfpa_sfpe_purser_2002}. Jin found visibility limits for occupants to be \SI{13}{\meter} if they were familiar and \SI{4}{\meter} if they are not familiar with the facilities~\cite{jin_1997}. Rasbash suggested a tenability criterion of \SI{10}{\meter} for safe escape \cite{rasbash_1975}. German building regulations refer to DIN 18009-2 \cite{din_din_18009_2} and vfdb guidelines \cite{vfdb_leitfaden}, specifying a minimum visibility of \SIrange{10}{20}{\meter}. Jin notes that wide variations in the proposed threshold values are probably due to differences in the geometry of the places and the composition of the group escaping from fire in the respective experimental investigations \cite{nfpa_sfpe_jin_2002}. Visibility threshold values are often expressed as optical density or extinction coefficient, correlating to visibility via the $C$ factor. For local assessment of ASET, such limits are more appropriate, as they are not related to a particular exit sign, which may not be visible at all. Furthermore, these values allow for a more accurate indication of other effects of the smoke, such as critical toxic exposure or eye irritation.


\section{Visibility maps}
\label{sec:visibility_maps}
\subsection{General concept}

The goal of visibility maps is to assess potential routes of egress and evacuation of building occupants, concerning the accessibility in the presence of fire induced smoke. In the context of a performance-based design, when numerical fire simulations are employed, visibility maps can become part of the building approval process. This makes it particularly important for the maps to be easily interpreted even by people without professional qualification.

Visibility maps are designed by indicating areas in the building floor plan which are safe or unsafe to walk on, with respect to sufficient visibility for safe egress. Significant differences to common procedures in practice arise both in considering the light obscuring effects of smoke and defining corresponding performance criteria. Visibility at a given location is calculated according to Jin's law (Eq. \ref{eq:jins_law}), although it is not considered to be a local quantity. Here, an effective or averaged value of the extinction coefficient is applied, which is computed in the line of sight between the viewpoint and an observed target. This takes particular account of a potentially inhomogeneous smoke distribution. Likewise, the required visibility automatically results from the distance between the viewpoint and the target. In accordance with Jin's law, reflecting or light emitting safety signs are considered as such targets. The location and characteristics of the $n$ exit signs along potential routes of egress are subsequently described as waypoints $W_k$ with $k \in \{1, \ldots, n\}$.

Visibility maps represent a Boolean matrix $M_{i,j}$, which labels the respective locations with physical Cartesian coordinates $X_i,Y_j$ as passable (1 or true) or non-passable (0 or false). Here, $i$ and $j$ represent the two-dimensional indices of the domain's agent floor cells. Only indices representing cells being accessible to persons (agents in the sense of simulation) and are not populated by architectural elements are considered in the evaluation. In this context, the extent of the matrix depends on the mesh-grid discretisation of the corresponding fire simulation. Nomenclature note: The visibility map $M_{i,j}$ is subsequently used with different additional indices, indicating if it is waypoint and/or time related, see the summarising Table~\ref{tab:variables} further on. According to Eq.~\ref{eq:available_vs_required_visibility}, the visibility map $M_{i,j,k}^t$, i.e. at a given time $t$ for the waypoint $W_k$, is obtained for all agent cells $i,j$ by matching the available visibility ($V_{i,j,k}^t$) against the required visibility ($L_{i,j,k}$). Here $L_{i,j,k}$ is the Euclidean distance between the agent cell $i,j$ and the waypoint $W_k$ and $V_{i,j,k}^t$ is the visibility based on the integrated extinction coefficient along the same visual axis. A key advantage of this approach is its independence from predefined performance criteria, as these emerge intrinsically from geometric features of the compartment.

\begin{equation}
M_{i,j,k}^t =
\begin{cases}
1, & \text{if } V_{i,j,k}^t \geq L_{i,j,k}\\
0, & \text{otherwise}
\end{cases}
\label{eq:available_vs_required_visibility}
\end{equation}

The time dependent visibility map $M_{i,j}^t$ is derived from the superposition of all $M_{i,j,k}^t$ using a logical OR operation, as described in Eq.~\ref{eq:vismap_logical_or}. The $M_{i,j,k}^t$ values are calculated independently for individual waypoints $W_k$. In this manner, all agent cells $i,j$ within the route of egress are marked passable if at least one exit sign is visible from there.

\begin{equation}
M_{i,j}^t = \bigvee_{k=1}^{n} M_{i,j,k}^t
\label{eq:vismap_logical_or}
\end{equation}

Thus, $M_{i,j}^t$ denotes a discrete temporal state of the visibility map, but can be further aggregated across a predefined set of time points $T$ of the fire simulation, see Eq.~\ref{eq:time_aggregated_vismap}. The time aggregated visibility map $M_{i,j}$ accordingly indicates whether the agent cells $i,j$ satisfy the visibility criterion according to Eq.~\ref{eq:available_vs_required_visibility} corresponding to the respective waypoints $W_k$ at each point in time $t \in T$.

\begin{equation}
M_{i,j} = \prod_{t \in T} M_{i,j}^t
\label{eq:time_aggregated_vismap}
\end{equation}

The above approach provides several advantages over the traditional assessment of visibility solely based on the local smoke density. First, the visibility factor $C$ in Jin's law was empirically derived by experimentally evaluating people's perception of exit signs in a smoke laden environment. Accordingly, any such signs must also be taken as a reference when applying the model in performance based design approaches based on numerical fire simulations. Furthermore, the approach also allows taking into account further aspects, such as the viewing angle or visual obstructions between the exit sign and the observer.

Generating visibility maps involves moderate computational effort and simply demands post-processing of simulation results from numerical field models such as FDS. For this purpose, spatially and temporally resolved information of smoke density or the extinction coefficient can be used. The analysis is conducted by means of the Python package FDS\emph{Vismap} \cite{fds_vismap}, which was developed by the authors and is made freely available. Here, the required data is extracted from horizontal FDS slice files (in the global $z$-plane), providing a transfer file format for visualising simulations results on planar cuts through the domain in the FDS post-processor software Smokeview \cite{smokeview_user_guide}. Employing the FDSReader, the data is transferred into two-dimensional NumPy arrays \cite{harris_numpy_2020} according to the dimension of the underlying mesh-grid. The FDSReader is a Python package for extracting multiple kind of results data from binary output files of FDS simulations.

For simplicity, the occupant's eye level is considered to be at the same height above the floor as the exit signs. Thus, the integration of the extinction coefficient along the visual axis is reduced to a two-dimensional level. Accordingly, the required extinction coefficient or smoke density is extracted from a single FDS slice data at a height of \SI{2}{\meter} above the floor level. This simplifies the actual three-dimensional configuration, as in general the perceiving eye and the perceived sign are not at the same height. Additionally, this reduces the amount of data needed, as no volumetric data of the smoke density is needed. Yet, these are only technical limitations and can be addressed in future implementations using the proposed methodology and techniques.

Multiple visibility maps $M_{i,j}^t$ may also contribute to the creation of ASET maps, like suggested in \cite{schroder_map_2020}. Following the ASET-RSET concept, ASET maps locally indicate the time when a specific performance criterion is first met or exceeded. For this purpose, local scalar quantities such as temperature, gas concentrations or visibility can be employed. ASET maps can likewise be created by first generating visibility maps for every point in time $t \in T$ within the observation period. In a ``first order'' ASET map, for a given set of points in time $T$, each agent cell $i,j$ is assigned the time $t$ at which the criterion according to Eq.~\ref{eq:available_vs_required_visibility} is not met for the first time (see Eq.~\ref{eq:aset_map}). ``First order'' in this context refers to the respective times only indicating when the corresponding exit sign is no longer visible. Higher-order ASET maps could potentially take into account the entire route of egress. This would imply manually or automatically identifying route decisions and tracking for every potential origin agent cell. Especially in crowded areas involving a large number of agents or in case of dynamic fire events, a supplementary evacuation simulation may be required for this purpose. A detailed elaboration of this concept will be the subject of future work.

\begin{equation}
\mathrm{ASET}_{i, j}=\min \left\{t \mid M_{i, j}^t=0, t \in T\right\}
\label{eq:aset_map}
\end{equation}

\begin{figure}[ht]
\centering
\includegraphics[width=0.8\textwidth]{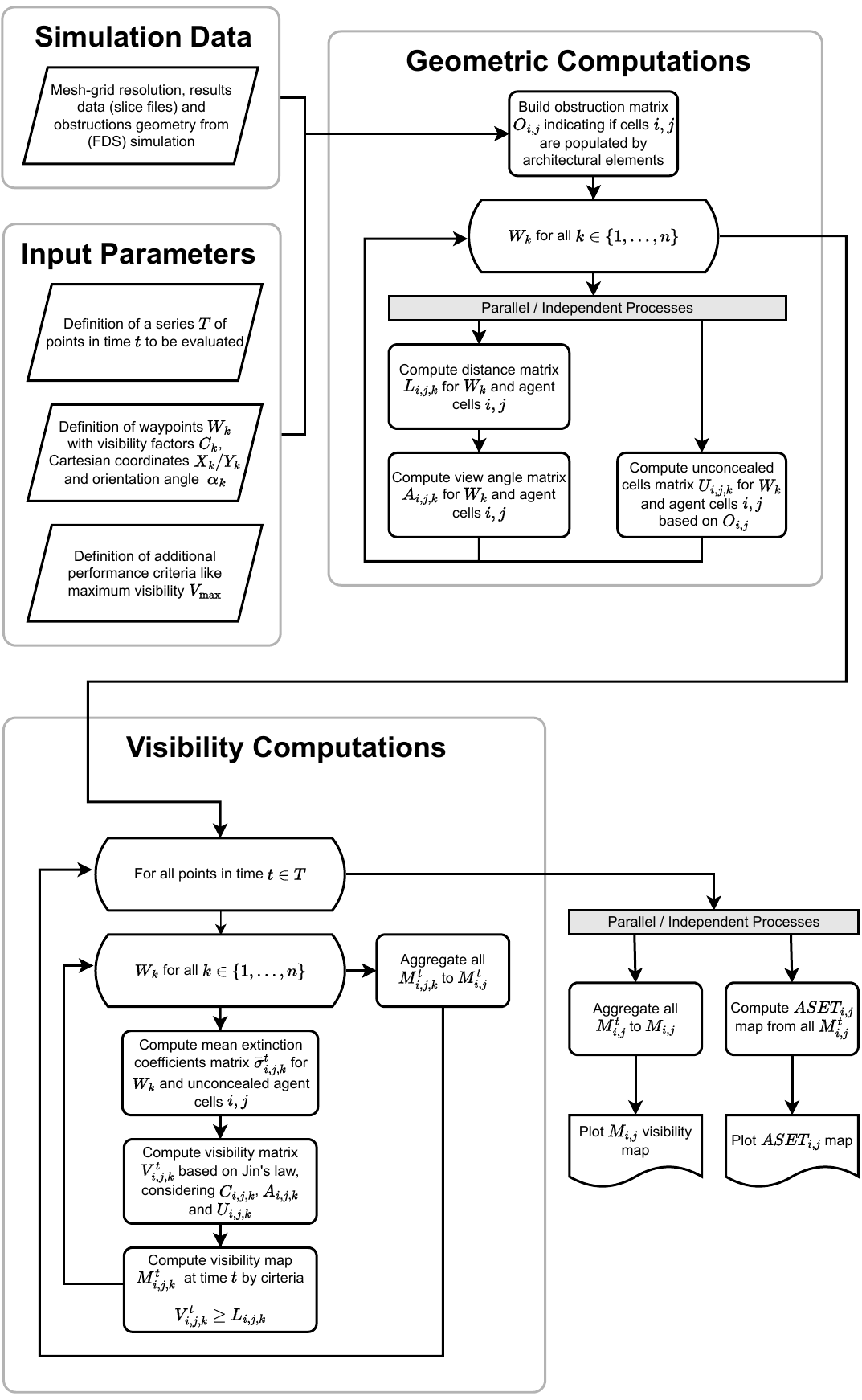}
\caption{Flow chart for the computation of visibility maps $M_{i,j}$ and $\mathrm{ASET}_{i,j}$ maps based on FDS simulation data. The algorithm can be subdivided into geometric computations and visibility computations. Both processes build on the simulation's boundary conditions and results, and on the defined waypoint parameters}
\label{fig:fdsvismap_flow_chart}
\end{figure}

The processing sequence to compute time aggregated visibility maps or ASET maps from the given input parameters is summarised in the program flowchart in Fig.~\ref{fig:fdsvismap_flow_chart}, while the following subsections provide further details about the procedure. Finally, it should be noted that processing data with high spatial (CFD grid) and temporal (number of evaluated points in time) resolution can take a considerable amount of time. Hence, it may be advisable to reduce the temporal increment outside critical periods of observation. This equally applies to the computation of time aggregated visibility maps and ASET maps.

\subsection{Model parameters and variables}
Waypoint parameters as well as the time points to be evaluated are defined detached from the imported fire simulation result data. A description of the required parameters and their respective data types and units is given in Table \ref{tab:parameters}. The angles and physical coordinates refer to a global Cartesian coordinate system with the same origin as the underlying fire simulation model. All coordinates are automatically interpreted as discrete indices $i,j$ according to the corresponding mesh-grid of the simulation.

\begin{table}[ht]
	\caption{Summarised description of the employed model parameters with their data types and units}
	\centering
	\begin{tabularx}{\textwidth}{l l l X}
		\toprule
		Parameter                & Type    & Unit  & Description \\
		\midrule
  		$T$              & List   & -     &  Set of points in time $t$ to be evaluated\\
		$W_{k}$              & Integer   & -     & Identifier of a waypoint\\
  		$X_k, Y_k$              & Float   & m     &  Physical Cartesian coordinates of the waypoint $W_k$ \\
        $C_{k}$              & Float   & -     & Visibility factor for the waypoint $W_k$, referring to the illumination type of the exit sign (usually $C=3$ for reflecting signs and $C=8$ for light emitting signs)\\
  		$\alpha_{k}$              & Float   & $^{\circ}$     & Rotation angle of the exit signs' observation normal at waypoint $W_k$ in the global coordinate system\\
		\bottomrule
	\end{tabularx}
	\label{tab:parameters}
\end{table}

All variables required to generate the visibility maps and ASET maps are described in Table \ref{tab:variables}. This data is either loaded from the simulation results or computed by means of simple mathematical or logical operations, taking into account the pre-defined waypoints. Most of these variables represent two-dimensional matrices of the same shape and extent as the fire simulation results data. 

\begin{table}[ht]
	\caption{Summarised description of the employed model variables with their data types and units}
	\centering
	\begin{tabularx}{\textwidth}{l l l X}
		\toprule
		Variable                & Type    & Unit  & Description \\
		\midrule
        $X_i, Y_j$              & Float   & m     &  Physical Cartesian coordinates of the agent cell $i,j$\\
    	$\theta_{k}$              & Float   & $^{\circ}$     & Viewing angle from the agent cell $i,j$ towards the observation normal of the exit sign at waypoint $W_k$ \\
		$O_{i,j}$              & Boolean   & - &  Indicating if cell $i,j$ is populated by an obstruction element (1 or True) or not (0 or False)\\
        $L_{i,j,k}$              & Float   & \SI{}{m} & Euclidean distance between agent cell $i,j$ and waypoint $W_k$\\
		$A_{i,j,k}$              & Float   & -- & Relative reduction of the emitted intensity as function of the view angle $\theta$ from agent cell $i,j$ towards the surface normal of an exit sign at waypoint $W_k$ \\
		$U_{i,j,k}$              & Boolean    & - & Indicating whether the agent cell $i,j$ is unconcealed (True or 1) or not (False or 0) from the waypoint $W_k$ \\
		$\rho_{\mathrm{s},k}^t(l)$              & Float    & \SI{}{\kilo\gram\per\meter} & Smoke density at distance $l$ from the agent cell $i,j$ towards the waypoint $W_k$ at time $t$ \\
		$\bar\sigma_{i,j,k}^t$              & Float    & \SI{}{\per\meter} & Integrated mean extinction coefficient along the line of sight between waypoint $W_k$ and agent cell $i,j$ at time $t$\\
		$V_{i,j,k}^t$              & Float   & \SI{}{\meter} & Visibility at agent cell $i,j$ at time $t$ with respect to the waypoint $W_k$ \\
  		$M_{i,j,k}^t$              & Boolean   & - & Indicating if the exit sign at waypoint $W_k$ is visible (1 or True) or not (0 or False) from agent cell $i,j$ at time $t$ \\
  		$M_{i,j}^t$              & Boolean   & - & Indicating if any exit sign is visible (1 or True) or not (0 or False) from agent cell $i,j$ at time $t$ \\
  		$M_{i,j}$              & Boolean   & - & Indicating if any exit sign is visible (1 or True) or not (0 or False) from agent cell $i,j$ at all points in time  $t \in T$ \\
  		$\mathrm{ASET}_{i,j}$              & Float   & \SI{}{\second} & Indicating the first point in time $t \in T$, the visibility criterion according to Eq.~\ref{eq:available_vs_required_visibility} is not satisfied at agent cell $i,j$\\
		\bottomrule
	\end{tabularx}
	\label{tab:variables}
\end{table}

A detailed explanation of the employed mathematical and physical models is provided below. The variables labelled with subscript $k$ are computed individually for each waypoint $W_k$. For reasons of computational efficiency, any computational operations will only be performed on agent cells $i,j$ within a radius corresponding to the maximum visibility $V_{\mathrm{max}}$ (usually \SI{30}{\meter}) around $W_k$.

\subsection*{Distance (Required Visibility)}
$L_{i,j,k}$ denotes the Euclidean distance according to Eq.~\ref{eq:euclidean_distance} between the waypoint $W_k$ with Cartesian coordinates $X_k, Y_k$ and the agent cell $i,j$ with Cartesian coordinates $X_i, Y_j$ in the evaluation level. It thus indicates the required visibility, for each location within the computational domain, with respect to the observed exit sign. The distance is only calculated in the observation plane and ignores any vertical height difference between the observer and the exit sign. In particular, for large distances, the resulting error can be considered negligible.

\begin{equation}
L_{i,j,k} = \sqrt{(X_i-X_k)^2+(Y_j-Y_k)^2}
\label{eq:euclidean_distance}
\end{equation}

\begin{figure}[!ht]
\centering
\begin{subfigure}{0.49\textwidth}
\centering
\includegraphics[width=\textwidth]{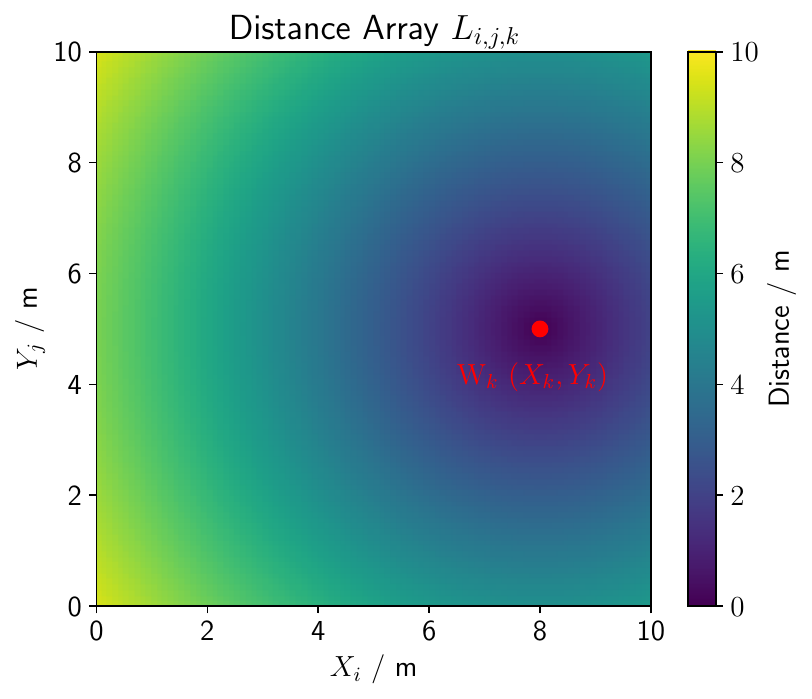}
\caption{}
\label{fig:distance_array}
\end{subfigure}
\hfill
\begin{subfigure}{0.49\textwidth}
\centering
\includegraphics[width=\textwidth]{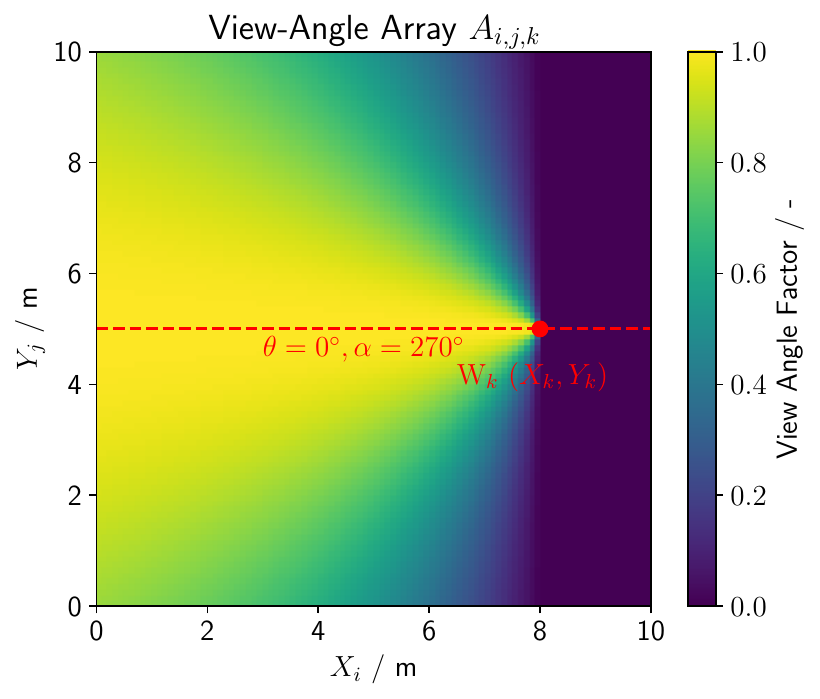}
\caption{}
\label{fig:angle_array}
\end{subfigure}
\caption{Exemplary visualisation of the distance matrix $L_{i,j,k}$ (b) and view-angle matrix $A_{i,j,k}$ (a) for the waypoint $W_k$ as a function of the Cartesian coordinates $X_i~/~Y_j$ of the agent cells $i,j$}
\end{figure}

An exemplary visualisation of the distance matrix $L_{i,j,k}$ is shown in Fig.~\ref{fig:distance_array}.

\subsection*{Viewing angle}
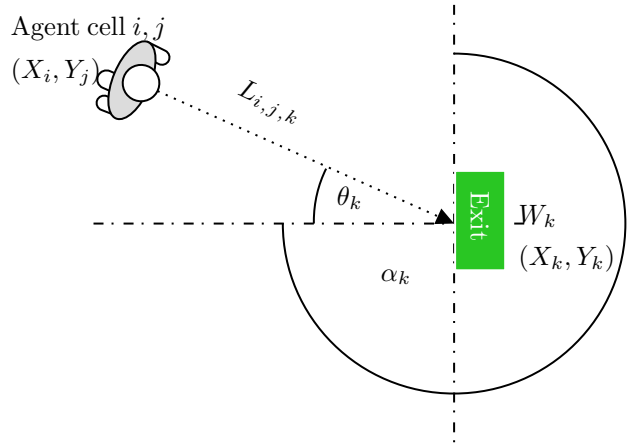
\begin{figure}[ht]
\centering

\tikzset{every picture/.style={line width=0.75pt}} 

\begin{tikzpicture}[x=0.75pt,y=0.75pt,yscale=-1,xscale=1]

\draw  [fill={rgb, 255:red, 255; green, 255; blue, 255 }  ,fill opacity=1 ] (251.89,35.1) .. controls (253.72,36.03) and (254.45,38.27) .. (253.52,40.1) -- (253.52,40.1) .. controls (252.59,41.93) and (250.36,42.66) .. (248.53,41.73) -- (239.77,37.28) .. controls (239.77,37.28) and (239.77,37.28) .. (239.77,37.28) -- (243.13,30.66) .. controls (243.13,30.66) and (243.13,30.66) .. (243.13,30.66) -- cycle ;
\draw   (219.91,64.83) .. controls (218,64.08) and (217.06,61.93) .. (217.81,60.02) -- (217.81,60.02) .. controls (218.55,58.11) and (220.71,57.16) .. (222.62,57.91) -- (233.06,61.99) .. controls (233.06,61.99) and (233.06,61.99) .. (233.06,61.99) -- (230.35,68.91) .. controls (230.35,68.91) and (230.35,68.91) .. (230.35,68.91) -- cycle ;
\draw   (220.62,53.71) .. controls (218.31,52.69) and (217.27,49.99) .. (218.29,47.69) -- (218.29,47.69) .. controls (219.32,45.38) and (222.01,44.34) .. (224.32,45.36) -- (229.34,47.59) .. controls (229.34,47.59) and (229.34,47.59) .. (229.34,47.59) -- (225.64,55.94) .. controls (225.64,55.94) and (225.64,55.94) .. (225.64,55.94) -- cycle ;
\draw  [draw opacity=0] (396,36.5) .. controls (396,36.5) and (396,36.5) .. (396,36.5) .. controls (396,36.5) and (396,36.5) .. (396,36.5) .. controls (443.22,36.5) and (481.5,74.78) .. (481.5,122) .. controls (481.5,169.22) and (443.22,207.5) .. (396,207.5) .. controls (348.78,207.5) and (310.5,169.22) .. (310.5,122) .. controls (310.5,121.92) and (310.5,121.84) .. (310.5,121.76) -- (396,122) -- cycle ; \draw   (396,36.5) .. controls (396,36.5) and (396,36.5) .. (396,36.5) .. controls (396,36.5) and (396,36.5) .. (396,36.5) .. controls (443.22,36.5) and (481.5,74.78) .. (481.5,122) .. controls (481.5,169.22) and (443.22,207.5) .. (396,207.5) .. controls (348.78,207.5) and (310.5,169.22) .. (310.5,122) .. controls (310.5,121.92) and (310.5,121.84) .. (310.5,121.76) ;  
\draw  [dash pattern={on 0.84pt off 2.51pt}]  (243.88,53.49) -- (267.96,64.34) -- (393.26,120.77) ;
\draw [shift={(396,122)}, rotate = 204.24] [fill={rgb, 255:red, 0; green, 0; blue, 0 }  ][line width=0.08]  [draw opacity=0] (8.93,-4.29) -- (0,0) -- (8.93,4.29) -- cycle    ;
\draw  [draw opacity=0] (326,122) .. controls (325.99,121.62) and (325.99,121.23) .. (325.99,120.85) .. controls (325.99,111.35) and (328.2,102.37) .. (332.12,94.39) -- (385.99,120.85) -- cycle ; \draw   (326,122) .. controls (325.99,121.62) and (325.99,121.23) .. (325.99,120.85) .. controls (325.99,111.35) and (328.2,102.37) .. (332.12,94.39) ;  
\draw  [dash pattern={on 3.75pt off 3pt on 0.75pt off 3.75pt}]  (216,122) -- (436,122) ;
\draw  [fill={rgb, 255:red, 220; green, 220; blue, 220 }  ,fill opacity=1 ] (243.13,30.66) .. controls (247.94,32.64) and (248.27,42.86) .. (243.88,53.49) .. controls (239.49,64.12) and (232.04,71.12) .. (227.23,69.14) .. controls (222.43,67.15) and (222.09,56.93) .. (226.49,46.3) .. controls (230.88,35.68) and (238.33,28.67) .. (243.13,30.66) -- cycle ;
\draw  [fill={rgb, 255:red, 255; green, 255; blue, 255 }  ,fill opacity=1 ] (234.32,43.97) .. controls (238.39,40.99) and (244.11,41.87) .. (247.09,45.94) .. controls (250.08,50.01) and (249.2,55.73) .. (245.13,58.71) .. controls (241.05,61.69) and (235.34,60.81) .. (232.35,56.74) .. controls (229.37,52.67) and (230.25,46.96) .. (234.32,43.97) -- cycle ;
\draw  [dash pattern={on 3.75pt off 3pt on 0.75pt off 3.75pt}]  (396,12) -- (396,232) ;

\draw (336,101.9) node [anchor=north west][inner sep=0.75pt]    {$\theta_k $};
\draw (358,143.4) node [anchor=north west][inner sep=0.75pt]    {$\alpha_k $};
\draw  [color={rgb, 255:red, 255; green, 255; blue, 255 }  ,draw opacity=1 ][fill={rgb, 255:red, 41; green, 198; blue, 35 }  ,fill opacity=1 ]  (396.5,95.5) -- (421.5,95.5) -- (421.5,145.5) -- (396.5,145.5) -- cycle  ;
\draw (409,120.5) node  [rotate=-90] [align=left] {\textcolor[rgb]{1,1,1}{{\fontfamily{helvet}\selectfont  \ Exit \ }}};
\draw (420,103.4) node [anchor=north west][inner sep=0.75pt]    {$ \begin{array}{l}
W_{k}\\
( X_{k} ,Y_{k})
\end{array}$};
\draw (290.8,46.4) node [anchor=north west][inner sep=0.75pt]  [rotate=-25.26]  {$L_{i}{}_{,}{}_{j}{}_{,}{}_{k}$};
\draw (166.6,9) node [anchor=north west][inner sep=0.75pt]    {$ \begin{array}{l}
\mathrm{Agent\ cell} \ i,j\\
( X_{i} ,Y_{j})
\end{array}$};

\end{tikzpicture}

\caption{The projection surface of the exit sign being observed changes according to the viewing angle $\theta$. $\theta$ can be described as a function of the orientation of the sign in the global $z$-plane, expressed by the rotation angle $\alpha_k$, and the viewer's position at the agent cell $i,j$.}
\label{fig:view_angle}
\end{figure}

The observation of signs at a certain angle, different from the normal, reduces its effective area of projection. Accordingly, the physical distance at which they can be perceived is reduced. Reflecting or light emitting exit signs can be considered as Lambertian radiators, so that the radiant intensity $I$ decreases by $\cos \theta$ with increasing view-angle $\theta$ to the sign \cite{din_din_iso_3864_1}. This effect is represented by $A_{i,j,k}$, considering that the sign can not be seen at view-angles $\theta \geq \ang{90}$ ($\cos \theta \leq 0$). Calculating $\cos \theta$ involves the dot product of the sign's surface normal and the observer's visual axis, as well as the observation distance $L_{i,j,k}$. Figure \ref{fig:view_angle} provides an illustration of the involved orientation angles in the global coordinate system. $A_{i,j,k}$ is computed according to Eq.~\ref{eq:view_angle}.

\begin{equation}
A_{i,j,k} = \max\left(0, \frac{\sin(\alpha_k)\cdot(X_i-X_k) + \cos(\alpha_k)\cdot(Y_j-Y_k)}{L_{i,j,k}}\right)
\label{eq:view_angle}
\end{equation}

$A_{i,j,k}$ reduces the maximum intensity relative to the observation normal of the respective exit sign as seen from cell $i,j$. Given the linear relationship between light transmission and the original intensity as stipulated by the Beer-Lambert law~\cite{1961.bouguer.beer_lambert_law}, this approach remains justified even when smoke is present in the observer's line of sight. An exemplary visualisation of the view-angle matrix $A_{i,j,k}$ is shown in Fig.~\ref{fig:angle_array}.

With the two-dimensional approach followed here, only the horizontal angle between the observer and the exit sign can be taken into account. However, since exit signs are usually located above the emergency exit doors and consequently, above the evaluation level, the approach entails a particular though minor degree of uncertainty. Incorporating the dependency of vertical viewing angles requires a spatial approach and will be the subject of future work.


\subsection*{Visual obstruction}
The visual obstruction of waypoints due to architectural elements is considered by the Boolean matrix $U_{i,j,k}$.  For this purpose, an auxiliary Boolean type matrix $O_{i,j}$ is first created in the shape of the simulation mesh-grid. In this, all cells are marked that are populated by a building component at the height of the observation plane. The obstruction coordinates on the FDS mesh-grid are automatically identified by the FDSReader from reading the respective Smokeview files. In this process, holes in the obstructions are considered by breaking the obstructions down into cuboid sub-obstructions.

\begin{figure}[ht]
\centering
\includegraphics[width=0.85\textwidth]{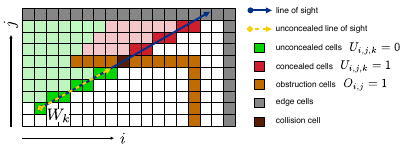}
\caption{A ray-casting algorithm is first employed to detect agent cells $i,j$ that are concealed by obstruction cells. The line of sight between the waypoint $\mathrm{W}_k$ and the edge cells is rasterised by the Bresenham's line algorithm or an advanced algorithm involving anti-aliasing. In the next step, the algorithm is employed to identify all cells that are considered in computing the average extinction coefficient $\bar\sigma_{i,j,k}$ between the waypoint $\mathrm{W_k}$ and all unconcealed agent cells $i,j$. The domain is discretised in the same shape as the numerical grid of the fire simulation.}
\label{fig:bresenham_line_algorithm_visibility}
\end{figure}

Detecting concealed cells, requires casting a ray from the waypoint to the individual cells to determine the collision coordinates with the obstruction cells. To obtain the number and location of the traversed cells until the collision point, the line of sight between the waypoint and the cells needs to be rasterised according to the respective mesh-grid. For this purpose, a simple drawing algorithm like the Bresenham's line algorithm \cite{bresenham_algorithm_1965} can be employed. The rasterized cells $p$ along each ray are described by the set $P_{\mathrm{cells}}$. In  $U_{i,j,k}$ all cells $p \in P_{\mathrm{cells}}$ are labelled as unconcealed (1, true) or concealed (0, false) when viewed from the waypoint $W_k$ depending on their location before or behind the first collision cell. Collision cells are defined as all cells along the rasterised line of sight that are populated by an obstruction in $O_{i,j}$. In order to reduce the number of cast rays and hence the required computational steps, only the edge cells of the mesh-grid are targeted in this process. An exemplary illustration of the ray-casting procedure is given in Fig.~\ref{fig:bresenham_line_algorithm_visibility}.

\begin{figure}[ht]
\centering
 \includegraphics[width=0.9\textwidth]{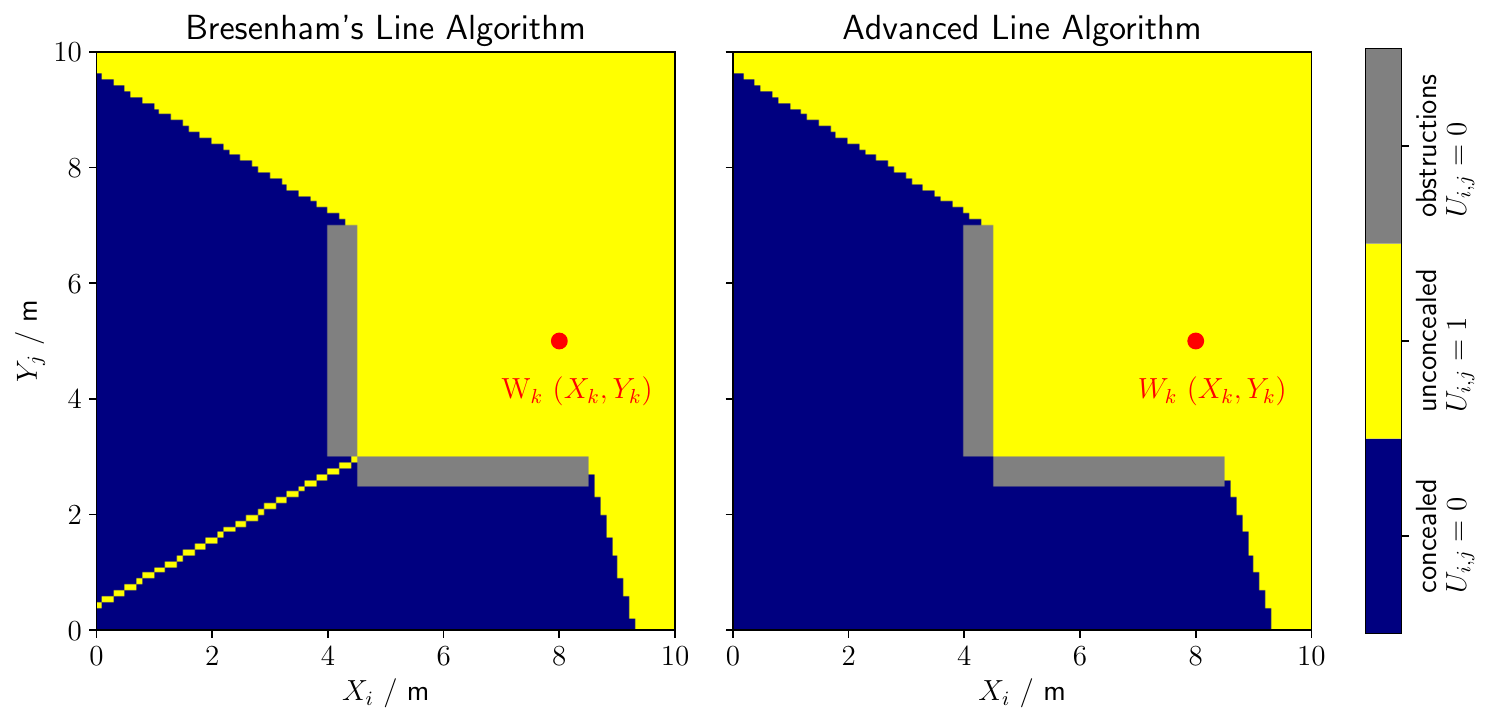}
\caption{In $U_{i,j,k}$ all agent cells $i,j$ are labelled as unconcealed (1, True) or concealed (0, False) when viewed from the waypoint $W_k$ depending on their location before or behind the first collision cell. When adjacent obstruction cells do not form an entirely closed barrier, a collision point may not occur with the rasterised line of sight. A more advanced ray-casting algorithm with a line thickness > 1 cell, e.g. using anti-aliasing, may then be applied to enforce an intersection.}
\label{fig:unconcealed_cells_array}
\end{figure}

Collision detection can become unreliable if adjacent obstruction cells do not form an entirely closed barrier. Due to the single row staircase structure created by the line algorithm, overlaps with the obstruction cells may not occur. Here an anti-aliasing technique, like introduced in \cite{wu_algorithm_1991} is employed to draw lines with a thickness of more than one cell. Such algorithm is computationally more complex than the Bresenham line algorithm, but is reasonable fast for the given application. The effect of applying an advanced line algorithm is illustrated by the visualisation of the matrix $U_{i,j,k}$ in Fig.~\ref{fig:unconcealed_cells_array}.

\subsection*{Available visibility}
Visibility according to Eq.~\ref{eq:jins_law} is primarily a function of the extinction coefficient $\sigma$ of the (fire generated) smoke between the observer and a visual target. In general, $\sigma$ can be expressed as a function of smoke mass density $\rho_{\mathrm{s}}$ and a mass-specific extinction coefficient $K_{\mathrm{m}}$. The latter indicates the contribution of the smoke to light extinction through absorption and scattering per unit mass. In compliance to the laws of classic electrodynamics $K_{\mathrm{m}}$ is a function of the wavelength $\lambda$ of the incident light. A uniform value of $K_m=\SI{8700}{\meter\per\kilo\gram}$ for red light at $\lambda=\SI{633}{\nano\meter}$ is widely adopted for proof of visibility in performance based fire safety design, as it is the FDS default value. It was derived from experimental investigations involving 29 different fuels with well ventilated flaming combustion \cite{2000.Mulholland}.

Assuming a constant value for $K_{\mathrm{m}}$, however, smoke density usually exhibits a temporal and spatial dependence. Given the general form of Beer Lambert's law, an effective or average extinction coefficient between the agent cell $i,j$ and the waypoint $W_k$ can be computed for penetrating an inhomogeneous medium at every time $t$ by Eq.~\ref{eq:mean_extinction_coefficient_general}. Here $\rho_{\mathrm{s}, k}^{t}(l)$ denotes the smoke density along the line of sight at the traversed distance $l$ towards the waypoint $W_k$ at time $t$. A similar approach to compute the mean extinction coefficient along the visual path from a camera to an exit sign crossing an inhomogeneous smoke layer is followed by Rinne et al. \cite{2007.Rinne}.

\begin{equation}
\bar\sigma_{i,j,k}^{t}=\frac{K_{\mathrm{m}}(\lambda) \int_0^{L_{i,j,k}}\rho_{\mathrm{s},k}^{t}(l) d l}{L_{i,j,k}}
\label{eq:mean_extinction_coefficient_general}
\end{equation}

Eq.~\ref{eq:mean_extinction_coefficient_general} can be  simplified to Eq.~\ref{eq:mean_extinction_coefficient_simplified}, assuming that the covered distance in all traversed cells $p \in P_{\mathrm{cells}}$ is identical, so that calculating the path increments can be omitted. Thus, $\sigma$ can be calculated directly as the arithmetic mean, depending on the number of traversed cells $\vert P_{\text {cells}}\vert$. This approximation significantly reduces the algorithm's complexity and the associated computational effort, while any loss in accuracy remains tolerable, given a sufficiently fine discretisation of the floor. Further optimisation is achieved by applying the algorithm exclusively to unconcealed cells, i.e. where $U_{i,j,k} \equiv 1$.

\begin{equation}
\bar\sigma_{i,j,k}^{t}=\frac{K_\mathrm{m}(\lambda)}{\vert P_{\text {cells}}\vert} \cdot \sum_{p\in P_{\mathrm{cells}}} \rho_{\mathrm{s},k,p}^{t}
\label{eq:mean_extinction_coefficient_simplified}
\end{equation}

The available visibility $V_{i,j,k}^{t}$ of all agent cells $i,j$ with respect to the waypoint $W_k$ can be calculated at any point in time $t$ according to Eq.~\ref{eq:available_visibility}. Here, the visual obstruction by architectural elements ($U_{i,j,k}$), the viewing angle ($A_{i,j,k}$) and the visibility factor $C_k$ are taken into account.
\begin{equation}
    V_{i,j,k}^{t} = \min \left( U_{i,j,k} \cdot A_{i,j,k} \cdot \frac{C_k}{\bar\sigma_{i,j,k}^{t}}, V_{\text{max}} \right)
    \label{eq:available_visibility}
\end{equation}

In performance based design, visibility is often limited to an arbitrary upper boundary value of $V_{max}=\SI{30}{\meter}$, since Jin's law is purely empirical and would imply an infinite visibility in the absence of smoke. Furthermore, the exit signs have a maximum visual distance even in a smoke-free environment.

\section{Application example}
\label{sec:application_example}
\subsection{Design Fire Scenario}
\label{sec:design_fire_scenario}

\begin{figure}[ht]
\centering
\includegraphics[width=0.7\textwidth]{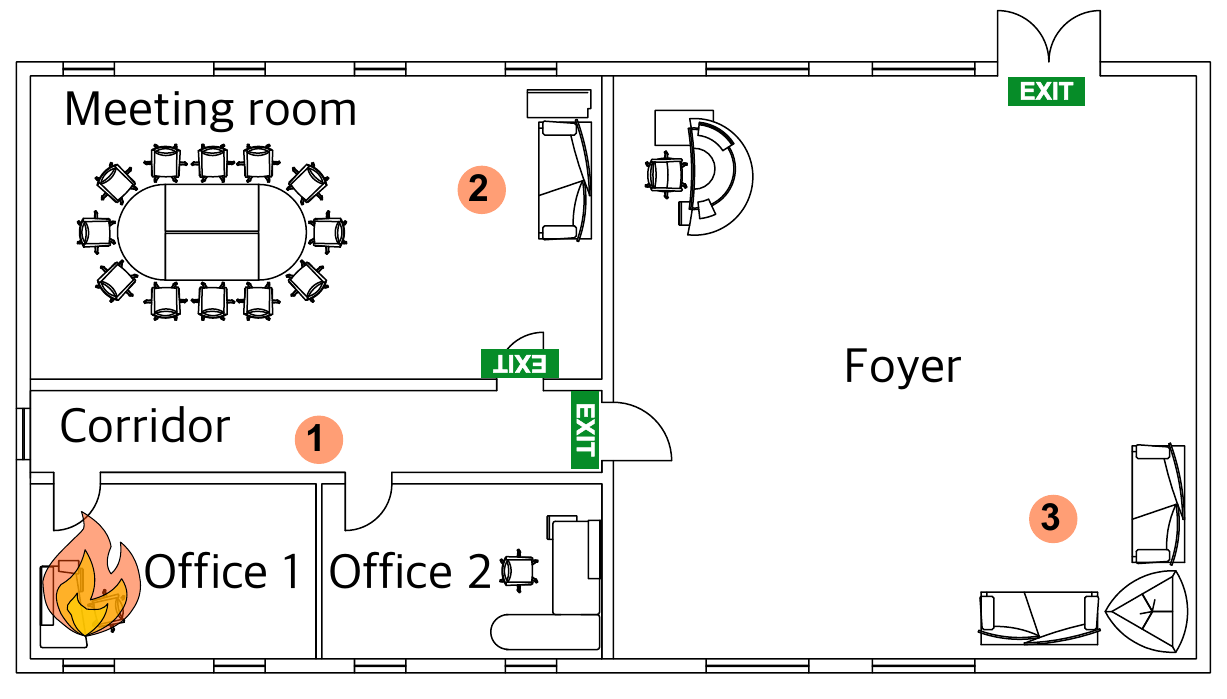}
\caption{Floor plan of the office facility  with dimensions of approximately $\SI{20}{\meter }~\times~\SI{10}{\meter}$. All rooms have a ceiling height of \SI{3}{\meter}, and the foyer has a ceiling height of \SI{4}{\meter}. The design fire is located in Office 1. Exit signs are located in the meeting room, corridor, and foyer as marked with orientation facing inside the respective rooms. Locations 1 – 3 (red circles) denote exemplary sampling points that are considered for a study on grid sensitivity.}
\label{fig:floorplan}
\end{figure}

\begin{figure}[ht]
\centering
\includegraphics[width=0.7\textwidth]{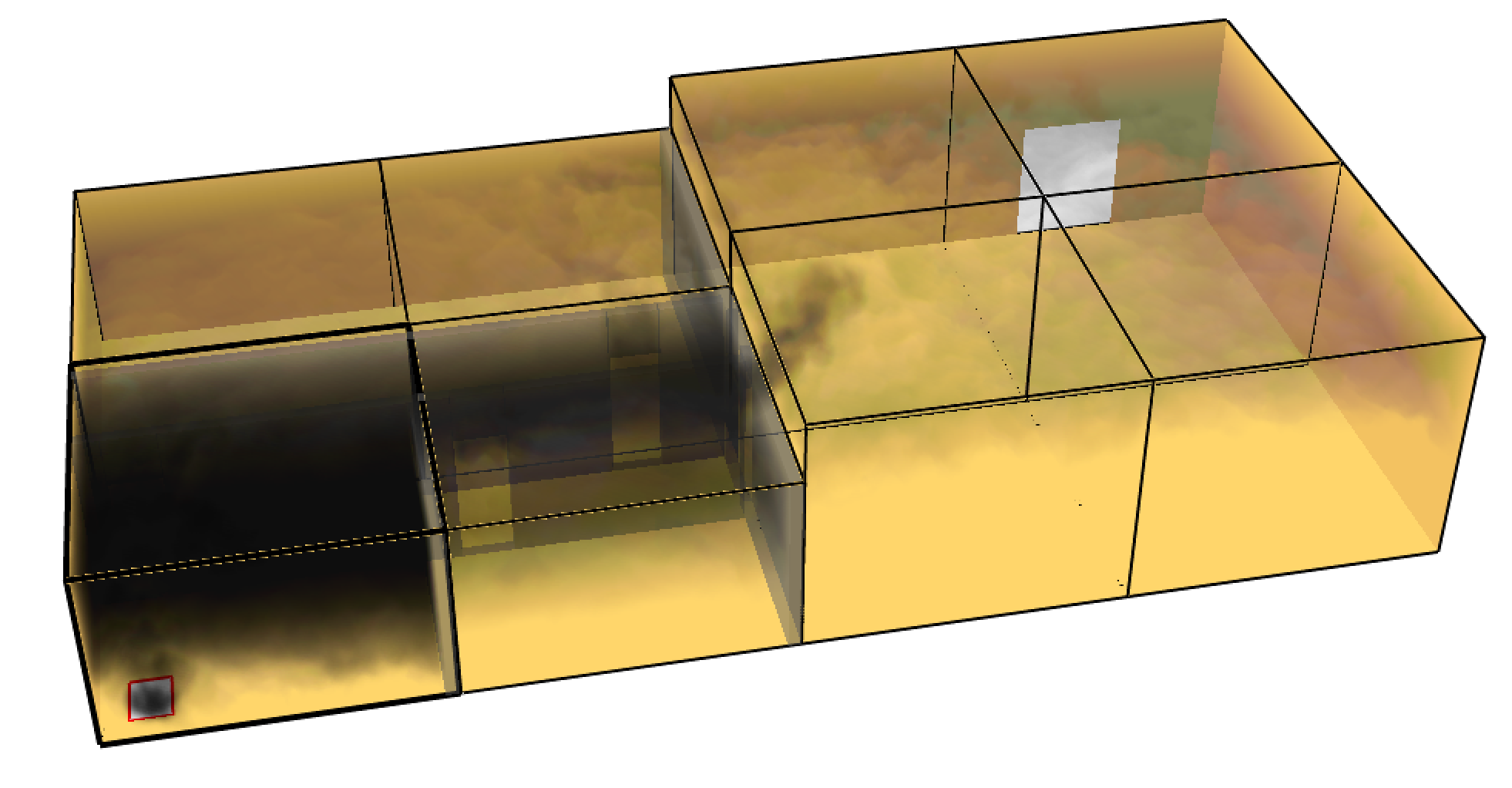}
\caption{Smokeview visualisation of the FDS simulation model showing the spread of fire generated smoke at $t=\SI{300}{\second}$. The computational domain is discretised by a uniform grid based on 8 meshes, each featuring a similar number of cells. The mesh boundaries represent the compartment enclosure, while the inner walls and doors are modelled as obstructions and holes.}
\label{fig:fds_model}
\end{figure}

The concept of visibility maps outlined above is demonstrated by means of a basic application example featuring a small office facility with dimensions of approximately $\SI{20}{\meter }~\times~\SI{10}{\meter}$, see Fig.~\ref{fig:floorplan}. All rooms have a ceiling height of \SI{3}{\meter}, and the foyer has a ceiling height of \SI{4}{\meter}. As shown in Fig.~\ref{fig:fds_model}, the buildings' geometry is mapped to a simple FDS model by discretising the computational domain with a uniform grid based on 8 meshes. The mesh boundaries form the compartment enclosure, while the internal components are mapped via FDS obstructions and holes. In the context of a grid sensitivity analysis, a total of 3 cases with cubic cells of \SI{5}{cm}, \SI{10}{cm} and \SI{20}{cm} edge lengths are examined. It should be noted that this study does not provide an isolated examination of the discretisation of the CFD model and the ray casting procedure. 

This example focuses on the assessment of unaided egress by the building's occupants. Accordingly, the design fire scenario is defined based on recommendations of the German VDI 6019 \cite{vdi_vdi_6019_1}\cite{vdi_vdi_6019_2} for an initial stage fire. The design fire assumes a linear increase of the heat release to $\dot{q}^{\prime\prime}_{\mathrm{max}}=\SI{100}{\kilo\watt}$ within \SI{300}{\second} followed by a constant progression on a constant burning area of $\SI{0.4}{\square\meter}$. The fuel is considered to be heptane, with a heat of combustion of $\Delta h_{\mathrm{c}} =\SI{46.6}{\mega\joule\per\kilo\gram}$ and a soot yield of $y_{\mathrm{s}}=0.037$, according to the SFPE Handbook of Fire Protection Engineering \cite{springer_sfpe_khan_2016}, assuming well ventilated combustion. The soot yield was adapted to $y_{\mathrm{s}}=0.04$ as a recommended value following ISO 13571 \cite{iso_iso_13571} for a typical plastic fire, applicable for performance-based design proof of visibility in case of fire.

The soot mass flux, introducing the combustion products in the computational domain according to the applied simple chemistry model of FDS, is given by Eq.~\ref{eq:soot_mass_flux}.

\begin{equation}
\dot{m}_{\mathrm{f,s}}^{\prime \prime}t=\frac{\dot{q}^{\prime \prime}t}{\Delta h_{\mathrm{c}}}\cdot y_{\mathrm{s}}
\label{eq:soot_mass_flux}
\end{equation}

In order to create the visibility maps, the spatio-temporal values of $\sigma_{i,j}^t$ are read from a single $z$-plane FDS slice-file with the quantity \texttt{Extinction Coefficient}. $\sigma_{i,j}^t$ is computed by FDS as a function of the smoke density $\rho_{\mathrm{s},i,j}$ and the mass specific extinction coefficient $K_{\mathrm{m}}$, which is set to the FDS default value of \SI{8700}{\meter\per\kilo\gram}.

The Python code for processing the FDS simulation data by using the FDSVismap package is given in Appendix B.

\subsection{Results and discussion}
\begin{figure}[ht]
\centering
\includegraphics[width=0.9\textwidth]{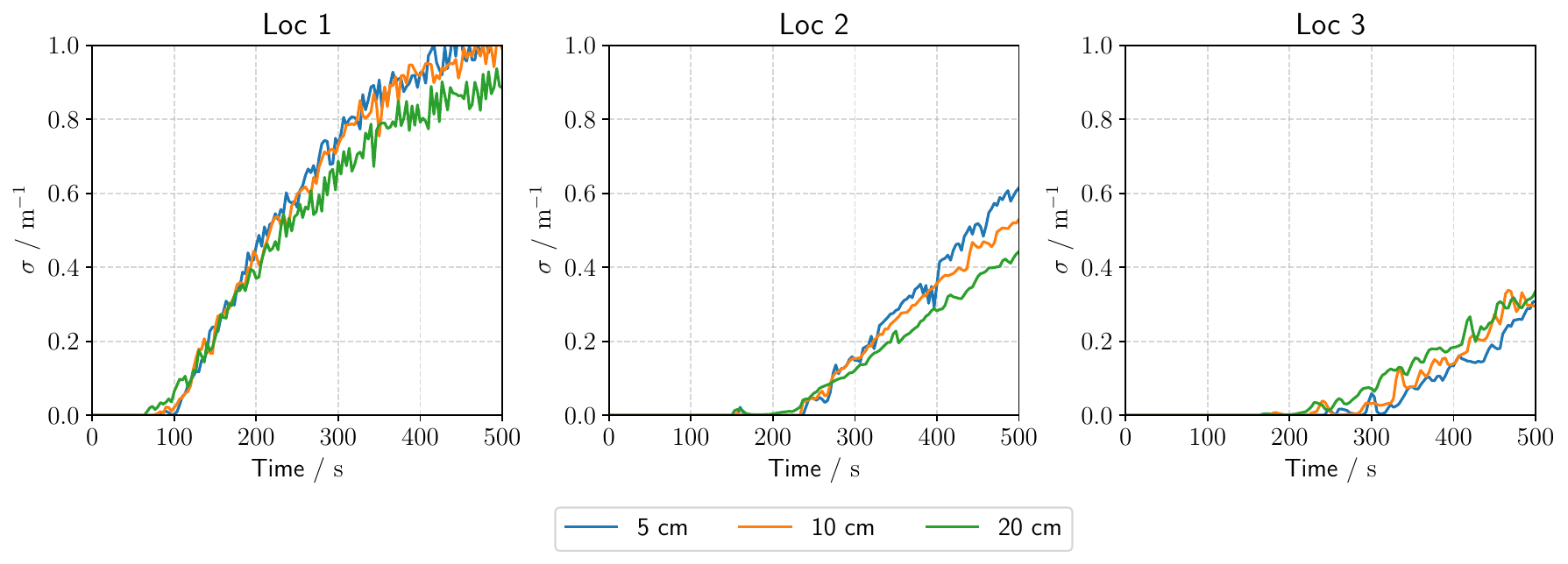}
\caption{Extinction coefficients as a function of time for three different locations according Fig.~\ref{fig:floorplan}. Values are given for three simulations with identical boundary conditions and geometry but different grid sizes with uniform cell sizes of \SI{5}{\centi\meter}, \SI{10}{\centi\meter} and \SI{20}{\centi\meter}.}
\label{fig:grid_sensitivity_analysis}
\end{figure}

\begin{figure}[ht]
\centering
\begin{subfigure}{\textwidth}
\centering
\includegraphics[width=0.7\textwidth]{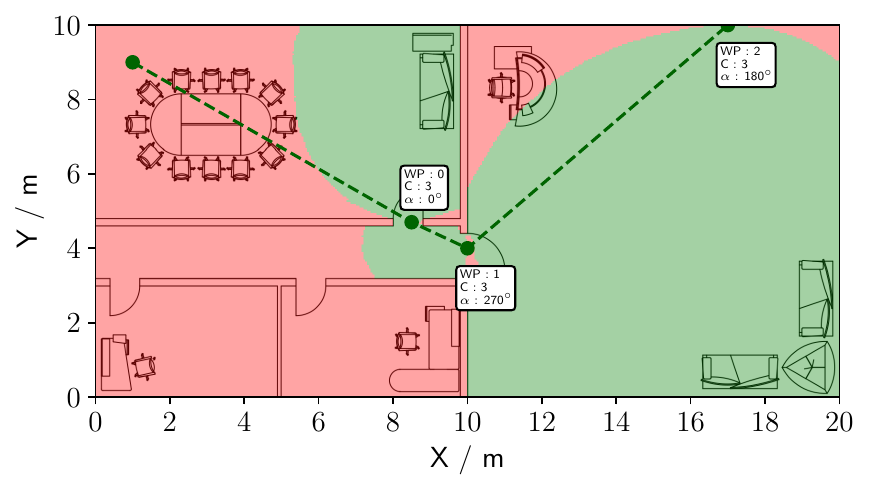}
\caption{Visibility factor $C=3$}
\label{fig:FDS_5cm_C3_400s}
\end{subfigure}
\hfill
\begin{subfigure}{\textwidth}
\centering
\includegraphics[width=0.7\textwidth]{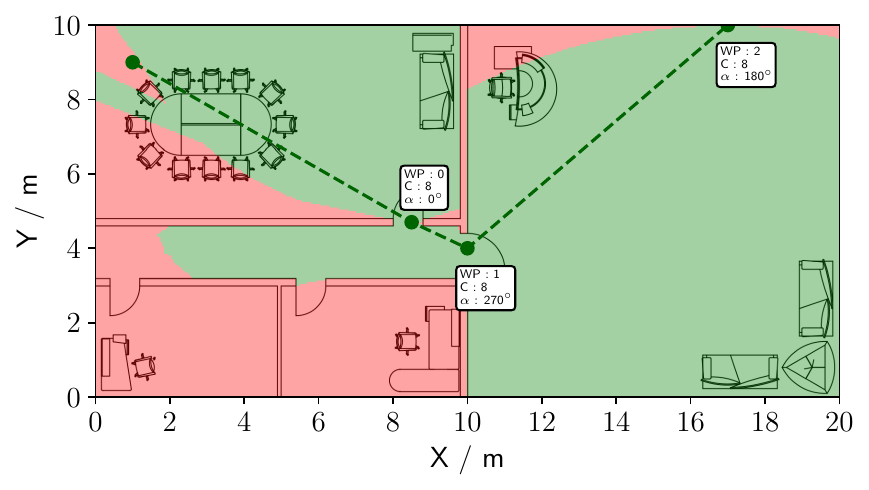}
\caption{Visibility factor $C=8$}
\label{fig:FDS_5cm_C8_400s}
\end{subfigure}
\caption{Visibility maps based on an FDS slice file with quantity \texttt{EXTINCTION COEFFICIENT} at $t=\SI{400}{\second}$ for a simulation with a uniform \SI{5}{\centi\meter} mesh-grid. Colours indicate if the visibility criterion according to Eq. \ref{eq:available_vs_required_visibility} is met (green) or not (red). The slice file is located \SI{2}{\meter} above the floor. All exit signs were assigned a visibility factor of $C=3$ (a) or $C=8$ (b)}
\end{figure}

\begin{figure}[ht]
\centering
\includegraphics[width=0.7\textwidth]{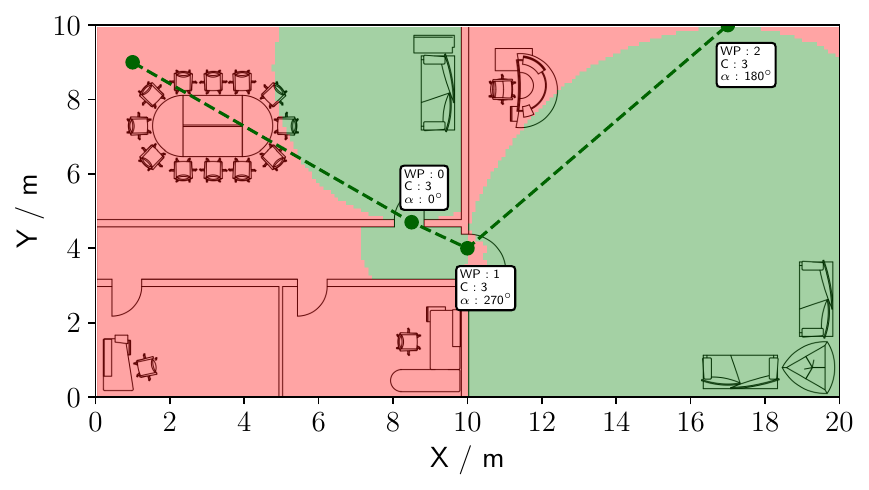}
\caption{Visibility map based on an FDS slice file with quantity \texttt{EXTINCTION COEFFICIENT} at $t=\SI{400}{\second}$ for a simulation with a uniform \SI{10}{\centi\meter} mesh-grid. Colours indicate if the visibility criterion according to Eq. \ref{eq:available_vs_required_visibility} is met (green) or not (red). The slice file is located \SI{2}{\meter} above the floor. All exit signs were assigned a visibility factor of $C=3$.}
\label{fig:Vismap_FDS_10cm_C3_400s}
\end{figure}

\begin{figure}[ht]
\centering
\includegraphics[width=0.7\textwidth]{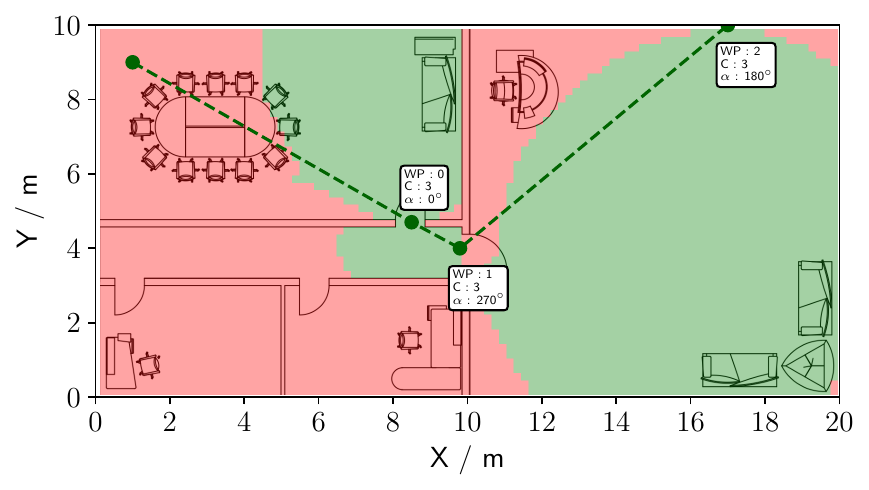}
\caption{Visibility map based on an FDS slice file with quantity \texttt{EXTINCTION COEFFICIENT} at $t=\SI{400}{\second}$ for a simulation with a uniform \SI{20}{\centi\meter} mesh-grid. Colours indicate if the visibility criterion according to Eq. \ref{eq:available_vs_required_visibility} is met (green) or not (red). The slice file is located \SI{2}{\meter} above the floor. All exit signs were assigned a visibility factor of $C=3$.}
\label{fig:Vismap_FDS_20cm_C3_400s}
\end{figure}

The influence of the model discretisation on the simulation results was analysed by means of a grid sensitivity analysis. Exemplary, the extinction coefficients at the locations Loc 1 - Loc 3 as (marked as red circles in Fig.~\ref{fig:floorplan}) were evaluated for three simulations with identical geometry and boundary conditions but different grid sizes (see Fig.~\ref{fig:grid_sensitivity_analysis}). The results of the three simulations generally show a consistent pattern, but exhibit quantitative discrepancies. At Loc 1, the \SI{5}{\centi\meter} and \SI{10}{\centi\meter} results converge, while the \SI{20}{\centi\meter} results lie somewhat below them. At Loc 2, all three simulations show the same rank but notable deviations that increase over time. No clear pattern is evident, resulting from the different grid resolutions at Loc 3, while the ranking of the extinction coefficient values reverses. Notable differences are also evident in the visibility maps generated from each simulation (see Fig.~\ref{fig:FDS_5cm_C3_400s} and Appendix A, Figs.~\ref{fig:Vismap_FDS_10cm_C3_400s} and \ref{fig:Vismap_FDS_20cm_C3_400s}). It is important to note that deviations resulting from different levels of discretisation may not solely be attributed to the simulation results, but also the conducted post-processing in generating the visibility maps. However, the influences from the FDS results appears to be dominant, as the ratio of the extinction coefficients according to Fig.~\ref{fig:grid_sensitivity_analysis} corresponds to the visibility range in the respective visibility maps.

Proof of visibility based on numerical fire simulation models is usually conducted at a certain point in time. In this context, the available time after the outbreak of the fire is determined until a route of egress is not passable any more and thus self-rescue is not possible. The evidence is provided if the available safe egress time (ASET) is longer than the required safe egress time (RSET). The RSET can be specified as a global performance criterion or derived from an individual assessment, e.g. by means of an evacuation simulation.

\begin{figure}[ht]
\centering
\includegraphics[width=0.7\textwidth]{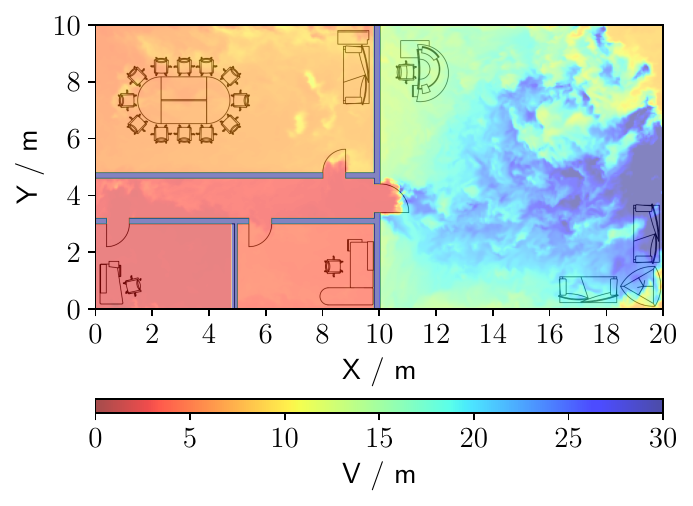}
\caption{Visualisation of the FDS Slice file with quantity \texttt{VISIBILITY} at $t=\SI{400}{\second}$ for a simulation with a uniform \SI{5}{\centi\meter} mesh-grid. The slice file is located \SI{2}{\meter} above the floor. The visibility factor was set to the FDS default value of $C=3$. Visibility is truncated at a maximum boundary of \SI{30}{\meter}.}
\label{fig:fds_visibility}
\end{figure}

Figure \ref{fig:fds_visibility} illustrates the challenges in interpreting visibility results as a direct output of the FDS simulation. It exemplary shows the local visibility according to Eq.~\ref{eq:jins_law} for a visibility factor of $C=3$, \SI{400}{\second} after the fire breakout. Across the corridor and both office units, visibility is below \SI{5}{\meter}, while in the meeting room it is generally about \SI{10}{\meter}. Due to a high level of turbulent flow, visibility in the foyer scatters from values of \SI{10}{\meter} to \SI{30}{\meter}. 

\begin{figure}[ht]
\centering
\includegraphics[width=0.7\textwidth]{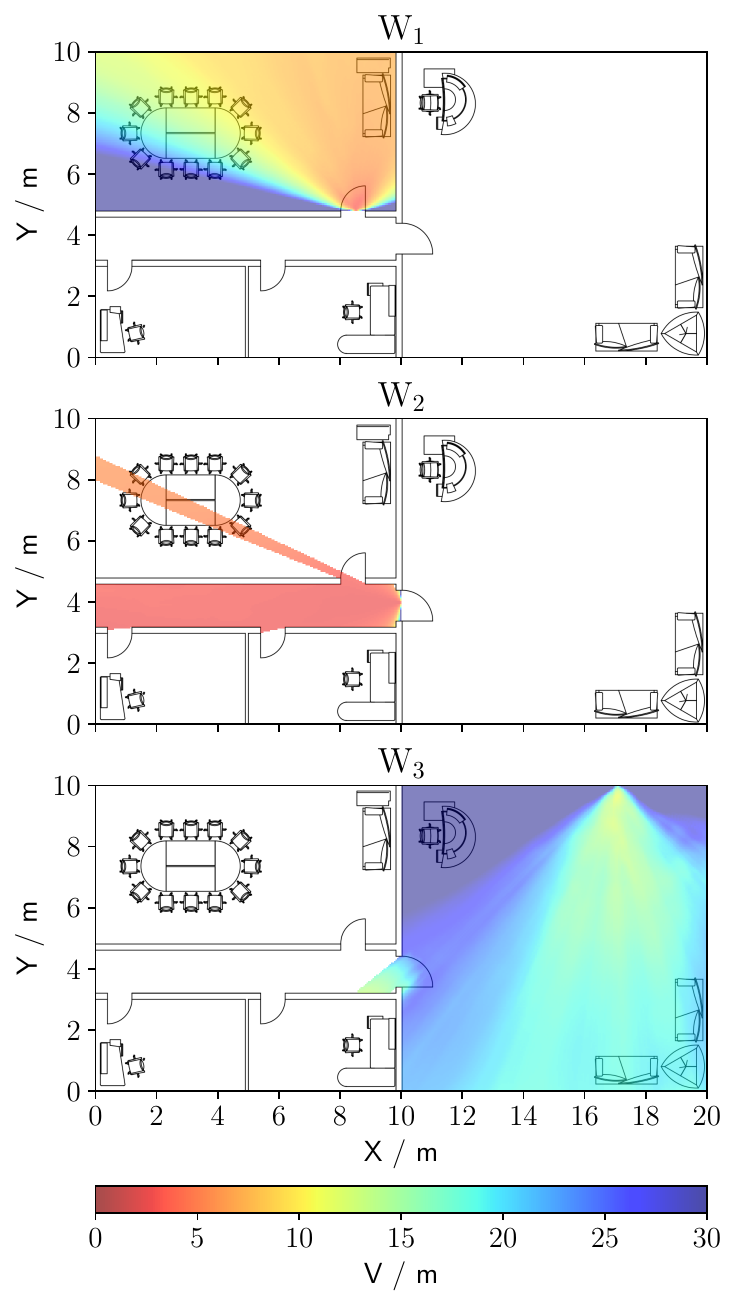}
\caption{Visualisation of the visibility matrices $V_{i,j,k}^t$ for the waypoints $\mathrm{W}_0-\mathrm{W}_2$ based on the FDS slice file with quantity \texttt{EXTINCTION COEFFICIENT} at $t=\SI{400}{\second}$ for a simulation with a uniform \SI{5}{\centi\meter} mesh-grid. The slice file is located \SI{2}{\meter} above the floor. The visibility factor was set to $C=3$ for all waypoints. Maximum visibility is truncated at \SI{30}{\meter}.}
\label{fig:fds_visibility_waypoints}
\end{figure}

Figure \ref{fig:fds_visibility} suggests that a lower threshold for minimum visibility of $\SI{10}{\meter}$ may be chosen. This way, the acceptable limits are met in the meeting room and in the foyer, while the limits are missed in the corridor. However, analysing and visualising the same scenario ($C_k=3$ for all $W_k$, $t=\SI{400}{s}$) by means of visibility maps results in a different assessment of the escape routes. In Fig.~\ref{fig:fds_visibility_waypoints}, the visibility matrices $V_{i,j,k}^t$ according to Eq.~\ref{eq:available_visibility} at the same time ($t=\SI{400}{\second}$) are shown for waypoints $W_1$ - $W_3$. The visibility map can subsequently be obtained by matching the available and required visibility (see Eq.~\ref{eq:available_vs_required_visibility}) and aggregating $M_{i,j,k}^t$ for all $W_k$ according to Eq.~\ref{eq:vismap_logical_or}. The visibility map (see Fig.~\ref{fig:FDS_5cm_C3_400s}) reveals that the corridor and foyer may be safely accessible by people escaping from the meeting room, while they might not see the exit sign inside the room. The primary distinction lies in the visibility map incorporating the actual boundary conditions and constraints throughout the designated escape route. The viewing angle significantly influences the perception of the exit sign, making it not visible in large parts of the meeting room. Conversely, for the considered escape scenario, a limited visibility is deemed acceptable in the corridor due to the shorter distance over which it is traversed.


Figure \ref{fig:FDS_5cm_C8_400s} shows a visibility map based on the same data but with a visibility factor of $C=8$ applied (recommended value for light emitting signs) for all exit signs. The map reveals that, to a large extent, conditions for safe egress in terms of visibility can be satisfied for occupants in the meeting room. However, it is important to note that this should not be considered as a comprehensive proof of safety. Due to potentially high local concentrations of smoke, individuals may be affected by other factors, such as eye irritation or reduced movement speed. Additional evidence may be required to ensure that tenable exposure to toxic smoke components is not exceeded.

Each of the visibility maps in Figs.~\ref{fig:FDS_5cm_C3_400s} and \ref{fig:FDS_5cm_C8_400s} present $M_{i,j}^t$ at discrete points in time, but with different values of $C$. An aggregated analysis spanning multiple temporal intervals may be advisable when the design fire scenario encompasses dynamic events that potentially affect smoke propagation. Such events could include opening and closing of doors and windows, as well as the operation of systems for natural or mechanical smoke extraction. Furthermore, automatic or manual fire suppression will consequentially affect the course of the fire. Furthermore, it is inevitably to examine all relevant design fire scenarios regarding potential fuels and locations.

\begin{figure}[ht]
\centering
\includegraphics[width=0.7\textwidth]{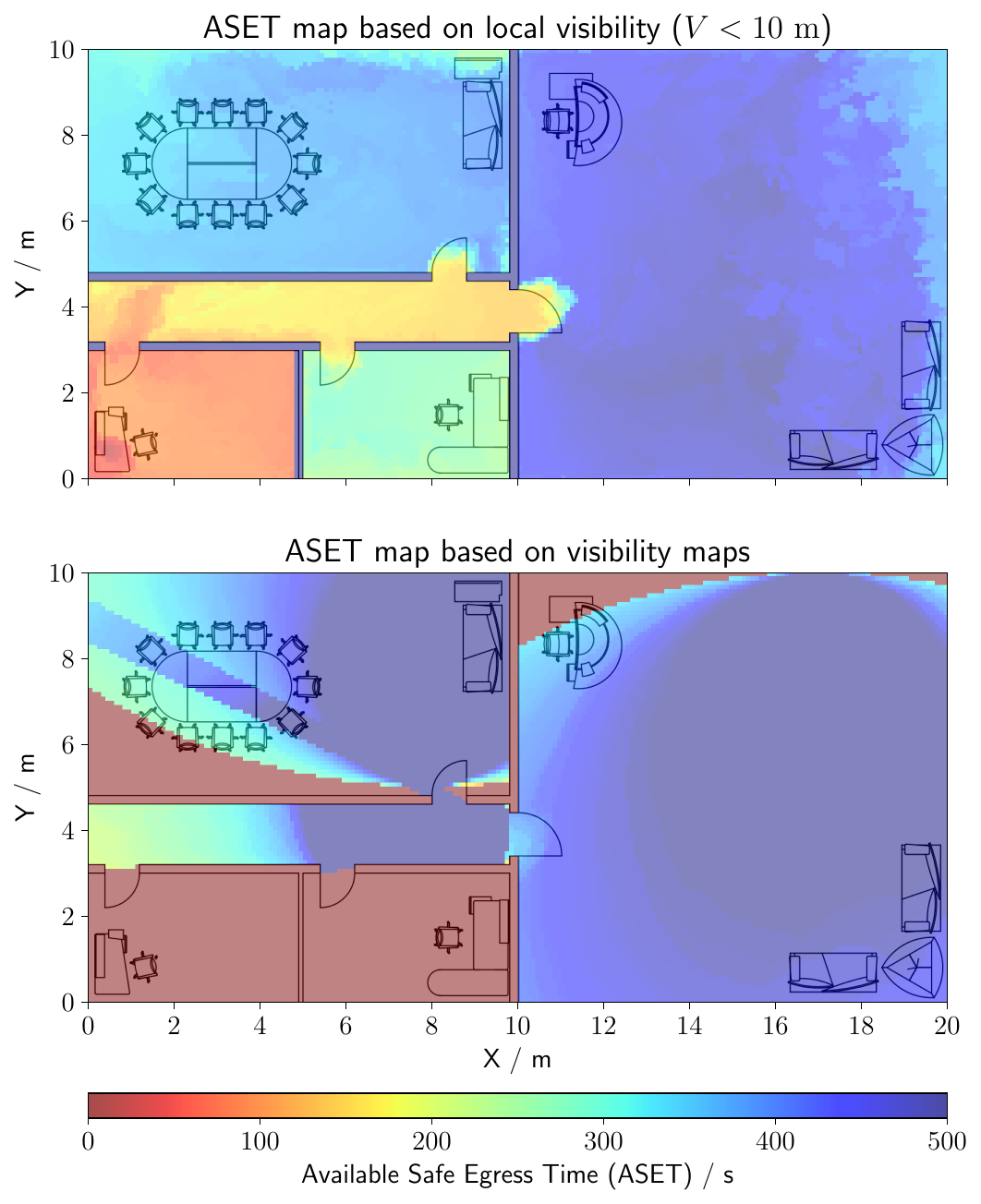}
\caption{ASET maps based on local visibility and visibility maps. Both maps rely on the same FDS slice file with quantity \texttt{EXTINCTION COEFFICIENT} for time series $T = \{ t \mid t = 0, 1, 2, ..., \SI{400}{\second}\}$. The simulation model features a uniform \SI{5}{\centi\meter} mesh-grid. The slice file is located \SI{2}{\meter} above the floor. All exit signs were assigned a visibility factor of $C=3$.}
\label{fig:local_vs_vismap_aset_map}
\end{figure}

Significant differences in the interpretation of local and path wise integrated visibility also become apparent in the representation as ASET maps (see Fig.~\ref{fig:local_vs_vismap_aset_map}) according to Eq.~\ref{eq:aset_map}. Both maps each are based on a visibility factor of $C=3$ for all exit signs. For the local ASET map, a lower visibility limit of \SI{10}{m} was defined. In case of the ASET map based on visibility maps, the required visibility is geometrically derived from the distances to the respective exit signs. Accordingly, both maps can only be compared on a qualitative level. The visibility map based ASET map indicates, certain areas of the floor plan do not have sufficient visibility even at time $t=0$. This primarily results from these areas being geometrically concealed from the exit signs. Hence, a critical interpretation by the user is required in such cases. Summarising, it can be noted that neither of the approaches can be generally considered as conservative. However, both the time aggregated visibility maps and the corresponding ASET map can reveal potential hazards (blind spots) that may not be recognised in a local assessment of visibility. Likewise, examining the local visibility can provide conclusions about potentially dangerous areas where occupants are likely to be exposed to an enhanced smoke concentration.

\section{Summary}
\label{sec:summary}

In this paper, an alternative approach is presented to assess visibility in case of a fire in the context of performance-based design. It adapts the commonly applied empirical law of Jin for use with inhomogeneous smoke distribution, as provided by CFD models. The method surpasses the traditional way of assessing local visibility, as commonly applied in fire safety engineering, by incorporating the location, orientation and type of exit signs (reflecting or light emitting) in the analysis.

The proposed approach enables a comprehensive assessment of egress routes, providing visibility maps or, in a more advanced form, ASET maps through post-processing of simulation data. It obviates the need for predefined visibility performance criteria, as these emerge naturally from the compartment's geometry and the distance between exit signs and observers. Additionally, the model automatically accounts for factors affecting the perception of exit signs, such as viewing angles and potential visual obstructions within the occupants' line of sight.

It was demonstrated that the commonly employed visualisation of visibility data based on the local smoke density shows a substantially different pattern compared to the introduced waypoint-based approach. The derived visibility maps allow a simple and distinct assessment of the simulation results. This can particularly contribute in identifying potential blind spots within a building design. Moreover, visibility maps are easy to interpret also by people without professional qualification and may therefore become part of the approval process for building designs. However, the straightforward nature of the data's visual representation should not prevent users from critically reviewing the applied model parameters and the resulting simulation results.

The introduced model was implemented as a Python package, which is made freely available: the Python package FDSVismap operates as a dedicated post-processor for data from FDS simulations, requiring only small additional effort from the user. The import of FDS data is handled with the open source Python module FDSReader. However, the data of any numerical fire model can generally be used to create visibility maps, if provided in a suitable transfer format. At the current state, the application generally requires moderate computational effort. However, especially for large simulations with dense mesh-grid discretisation involving multiple time steps to be evaluated, generating visibility maps can be highly time-consuming. An optimisation of the model is the subject of future work.


\section{Outlook}
\label{sec:outlook}
A significant focus of future work will be directed towards optimising the algorithms of the Python implemented FDSVismap model. This especially involves vectorisation of computational operations, particularly concerning ray casting and data caching, to avoid redundant computations. Additionally, implementing advanced techniques like quad tree algorithms could significantly improve collision detection with obstruction cells.

Future iterations of the model aim to facilitate a spatial analysis, allowing exit signs to be considered that are located beyond the horizontal observation plane. This would allow for an enhanced assessment of escape scenarios in more complex, multi-storey building environments. At this stage, ASET maps, generated on the basis of the visibility maps, still need to be interpreted in the context of a particular evacuation scenario. However, higher-order ASET maps may encompass the entire route of egress, requiring (automatic) identification of location based escape route decisions, potentially integrating data from pedestrian dynamics simulations. 

Regardless of the presented approach, it is important to point out that the frequently used visibility model according to Jin needs to be fundamentally revised. Notably, the existing model  does not adequately account for the absorbing and scattering behaviour of different kinds of smoke that might emerge from flaming or smouldering combustion. The model also provides limited scope in considering environmental effects, such as ambient light, which can significantly affect the contrast and thus the visibility of exit signs. To overcome these limitations, extensive experimental investigations and model developments will be required in the future.

\newpage
\section*{Appendix}
\label{sec:Appendix}
\subsection*{Appendix A -- Grid sensitivity analysis for visibility maps based on different FDS mesh-grid resolutions}
\label{sec:Appendix_a}
\FloatBarrier

\FloatBarrier
\subsection*{Appendix B -- Python Code for the example shown in section \ref{sec:application_example}}
\label{sec:Appendix_b}

\begin{lstlisting}[language=Python]
# Import FDSVismap package
from fdsvismap import VisMap

# Set path for FDS simulation directory
sim_dir = 'fds_data'

# Create instance of VisMap class with maximum value for visibility
vis = VisMap(sim_dir, max_vis=30)

# Add background image
bg_img = 'floorplan.png'
vis.add_background_image(bg_img)

# Set starpoint and waypoints along escape route
# Waypoints have parameters X_k, Y_k, C_k and alpha_k
vis.set_start_point(1, 1)
vis.set_waypoint(8, 5.5, 3, 0)
vis.set_waypoint(10, 6, 3, 270)
vis.set_waypoint(17, 0, 5, 180)

# Set time points when the simulation should be evaluated
vis.set_times([400])

# Do the required calculations to create the visibility maps
vis.compute_all()

# Plot time aggregated visibility map
vis.plot_time_agg_vismap()

\end{lstlisting}

\section*{Acknoledgements}
\label{sec:acknoledgements}
This work was largely funded by the NextVIS project of the German Research Foundation (Deutsche Forschungsgemeinschaft -- DFG) with the grant number 465392452. The authors gratefully acknowledge the financial support of the German Federal Ministry of Education and Research. The extensive computations of FDS simulations were performed largely on the high-performance computer system funded as part of the CoBra project with the grant number 13N15497.

\section*{Authorship Contribution Statement, following \citep{Allen2019}}

\textbf{Kristian Börger:} conceptualisation, formal analysis, investigation, methodology, software, validation, visualisation, writing -- original draft preparation, writing -- review and editing

\textbf{Alexander Belt:} conceptualisation, validation, writing -- review and editing

\textbf{Lukas Arnold:} conceptualisation, methodology, project administration, resources, supervision, writing -- review and editing, funding acquisition

\bibliographystyle{unsrtnat}
\bibliography{references}  






\end{document}